\documentclass[a4paper,12pt]{article}

\usepackage{amssymb,amsmath,mathbbol,mathrsfs}
\usepackage[usenames,dvipsnames]{color}
\usepackage{hyperref}
\usepackage{xcolor}
\usepackage{stmaryrd}
\usepackage{authblk}
\usepackage{framed}
\usepackage{empheq}
\usepackage{slashed}
\usepackage{hyphenat}
\usepackage{cite}
\usepackage{apacite}
\usepackage{natbib}
\usepackage{dsfont}
\usepackage{chngcntr}
\usepackage{enumerate}

\usepackage{marginnote}    

\usepackage[left=.8in,right=.8in,top=.8in,bottom=.8in]{geometry}                  

\usepackage{sectsty} 
\allsectionsfont{\sffamily\mdseries\upshape} 
\usepackage{tocloft}

\makeatletter
\renewenvironment{abstract}{%
    \if@twocolumn
      \section*{\abstractname}%
    \else 
      \begin{center}%
        {\bfseries\sffamily\abstractname\vspace{\z@}}
      \end{center}%
      \quotation
    \fi}
    {\if@twocolumn\else\endquotation\fi}
\makeatother

\numberwithin{equation}{section}
5

\hypersetup{
	colorlinks=true,         
	linkcolor=MidnightBlue,          
	citecolor=BrickRed,        
	urlcolor=MidnightBlue            
}

\newcommand{\be}{\begin{equation}}
\newcommand{\ee}{\end{equation}}

\renewcommand{\d}{{\mathrm{d}}}
\newcommand{\bbS}{\mathbb{S}}
\newcommand{\D}{{\mathrm{D}}}

\renewcommand{\bar}{\overline}

\renewcommand{\hat}{\widehat}
\newcommand{\RR}{\mathds{R}}

\newcommand{\cint}{{\int\kern-.87em{<}}}
\newcommand{\sint}{{\int\kern-.75em{\sim}}}
\newcommand{\fint}{{\int\kern-1.00em{\int}}}

\let\oldmarginpar\marginpar
\renewcommand\marginpar[1]{\oldmarginpar{\color{red}\raggedright\footnotesize #1}}

\title{On the analogies between gravitational and electromagnetic radiative energy}
\date{24th of January, 2023}
\author{Henrique Gomes\footnote{University of Oxford, Oriel College, OX1 4EW, United Kingdom; \href{mailto:gomes.ha@gmail.com}{gomes.ha@gmail.com}}~ and Carlo Rovelli\footnote{Aix-Marseille University, Centre de Physique Th\'eorique, Marseille, France; \href{mailto:rovelli@cpt.univ-mrs.fr}{rovelli@cpt.univ-mrs.fr }  }  }
\begin{document}
\maketitle
\begin{abstract} 
 We give a conceptual exposition of aspects of gravitational radiation, especially in relation to energy. Our motive for doing so is that the strong analogies with electromagnetic radiation seem not to be widely enough appreciated. In particular, we reply to some recent papers in the philosophy of physics literature that   seem to deny that gravitational waves carry energy.

Our argument is based on two points: (i)  that for both electromagnetism and gravity, in the presence of material sources,  radiation is an effective concept, unambiguously emerging only in certain regimes or solutions of the theory; and (ii) similarly, energy conservation is only unambiguous in certain regimes or solutions of general relativity. Crucially, the domain of (i), in which radiation is meaningful, has a significant overlap with the domain of (ii), in which energy conservation is meaningful.   Conceptually, the overlap of regimes is no coincidence: the long-standing question about the existence of gravitational waves was  settled  precisely by finding a consistent way to articulate their energy and momentum. 
  \end{abstract}

\section{Introduction}

Heuristically, gravitational waves are propagating ripples in the fabric of spacetime; ripples  that   we can now detect as originating in some of the most energetic events in the universe, such as the merger of two black holes.   In recent years, instruments such as LIGO (the Laser Interferometer Gravitational-Wave Observatory) have made it possible to measure these remarkably elusive waves. Since the first direct detection 
in 2015,  there have been numerous detections, 
 including the merger of two neutron stars and the collision of two black holes. It is not hyperbole to say we have developed a new type of sensor 
 with which to  observe the cosmos.
 
 But in earlier decades, gravitational waves were controversial. The theory governing their 
behaviour 
 had a turbulent origin.   Proposed by Albert Einstein soon after the discovery of general relativity, gravitational waves were shortly thereafter confused with mere artefacts  of a bad coordinate choice; and so their existence was denied. But Einstein himself was thereafter convinced  they were not just  coordinate artefacts, and so came to realise the waves are real \citep{Kennefick2007}. But   these controversies are now long  past. The work of Penrose, Bondi, Metzner, van der Burg and Sachs in the 1960s provided an invariant formulation with which to describe gravitational radiation at asymptotic distances from their sources. Crucially, gravitational waves were shown to carry energy in an unambiguous, fully covariant manner. 

 But scepticism about whether these waves carry energy lingers on in some  of the philosophy of  physics literature  (see e.g. \cite{Hoefer_grav, Lam_grav, Duerr_grav, Fletcher_grav}). 
Their sceptical arguments have two main foci: they attempt to (i)    show germane dissimilarities between gravitational and other types of physical interactions, and to (ii) weaken the significance of asymptotic conservation laws.   In the course of the paper, we will---one might say, unduly---focus on \cite{Duerr_grav}. But there is a simple reason: it is the most recent and complete illustration of (i) and (ii).

 Thus we plan to review the established work from the 1960s (and some later developments), including the analogies between  gravitational radiation and electromagnetic radiation, with its much better understood energy transfer: and thereby reply to these sceptical arguments.

There is general philosophical debate that we do not intend to here address about (i) the nature of idealizations and approximations and (ii) the nature of the contrast between emergent or effective regimes, and fundamentals. That is why we  use  electromagnetic radiation as a foil: reasonable views on (i) and (ii) should apply equally to electromagnetic and gravitational radiation. 

  In the rest of this Section, we briefly introduce the paper's themes, and give a prospectus.
 
One argument against gravitational radiation carrying energy is that it can be sourced by objects that are following geodesics, and are thus `force-free'. But the generation of gravitational waves depends on  the quadrupole moment of the source's motion: that is a statement about relative motion between different bodies, or parts of a body. Whether or not the individual bodies or parts of a body follow `force-free' motion, that motion produces real tension, and thus work, within an extended body.  Thus, being force-free is compatible with the emission or absorption of energy. 

 We can  see this very clearly in the electromagnetic analogy. Take  the question of whether a freely falling charge radiates or not. This  was satisfactorily answered in the 1960s by Rohrlich and Fulton \citep{Fulton1960}. Their  answer was that the freely falling particle can be taken to  radiate, for us, in Earth's frame, but not for a local, comoving observer. 
 
 The reason this answer is   (perhaps surprisingly) not in conflict with the equivalence principle is that, generically, we can't unambiguously extract the wave part of the electromagnetic field from its electro-magneto-static part, within spatial regions that are close to the source of the wave. Even in a Minkowski spacetime, this extraction is only unambiguous at large distances:  the so called \emph{wave-zone} (see appendix \ref{app:EM}).

And  there is a very analogous picture for gravitational waves. In a weak-field approximation,  it is in the wave-zone of the source that we can only unambiguously discern the part of the linearised field that is radiative. To accomplish something similar in full general relativity, we also  need to invoke something akin to the wave-zone: asymptotic infinity. Generically, it is only asymptotically that we have an objective split between the ``Newtonian" (or Poisson, or Coulombic) part, and  the radiative part,  of the field. 

At this point it is worthwhle to disambiguate between two different  meanings of `radiation'. For one could ascribe to `radiation'  a very weak, and even vague, meaning:  the propagating effect of a given change in initial conditions (an effect that by all accounts  travels within the light-cone).\footnote{ Both general relativity and electromagnetism respect relativistic causality, which means that, given two solutions of the equations of motion that agree on a proper subset of a Cauchy initial slice $\Sigma$, i.e. that agree on $\Sigma_0\subset \Sigma$, will agree on $D^+(\Sigma_0)$  (the domain of dependence  of $\Sigma_0$). See \cite[Appendix III, Theorem 2.15]{choquet2008} and \cite[Appendix B]{Landsman_GR} for a proof, which is based on the fact that both the Einstein and the Maxwell equations are quasi-linear hyperbolic equations. By the same token, differences in initial data constrained to a subset will evolve into differences within the domain of dependence of that subset. There is of course the tricky question of how to actually build solutions that match in some subset but differ elsewhere, which we discussed in footnote \ref{ftnt:gluing}.} 

But the weak meaning of radiation is not the one that features in discussions of gravitational waves in the weak-field limit, and they are not the sense we are mostly concerned with here.  The stronger meaning of radiation that we are concerned with here is similar to the meaning it has in flat spacetime: waves should  propagate along null directions, and have transversal polarizations, oscillatory behavior, and, when sourced by compact sources,  the correct fall-off behavior. 

Thus we see that, properly understood,  `radiation' is a derivative concept; it is a property of the field that in limited regimes emerges out of the fundamental ontology of each theory. In both the gravitational and electromagnetic cases, generically,   it is only only at asymptotic distances that we can \emph{unambiguously} distinguish radiative and `Coulombic' components. And in both cases there are similarly special circumstances---that are again similar in the electromagnetic and gravitational cases---in which we can distinguish these derivative or emergent properties everywhere. 
  
  In sum, singling out the `waving' part of the field  is  subtle {\em both} for electromagnetism and gravity,  and in similar ways. 
 We will thus argue that  dissimilarities between electromagnetic and gravitational waves are irrelevant to the issue at hand. That is,  we can happily admit that spacetime curvature and matter fields are  fundamentally distinct---``marble and wood'', as Einstein called them.\footnote{ The labels are usually taken to mean that the geometric side of the equations (the Einstein tensor and cosmological constant) was smooth and pristine---like marble---whereas the right-hand side of the Einstein equations (the energy momentum tensor),  was of a different, knottier, or rougher nature, `like low-grade wood'. But  as discussed in \cite[Sec 3]{Lehmkuhl_genesis}, it is not quite what Einstein meant: \begin{quote}
 Einstein seeing the left-hand
side of the Einstein equations as fine marble and the right-hand
side as low-grade wood has nothing to do with geometry. It
is about quanta. He believed that the left-hand side of the Einstein
equations gave an accurate picture of the gravitational field, but
that the right-hand side of the equations did not give an accurate
picture of matter, for it does not account for the quantum features
of matter. It is only a docking station for results of theories like
classical hydrodynamics and electrodynamics, which do not do
justice to the quantum nature of matter either. Thus,
$T_{\mu\nu}$ 
in GR is a
place-holder for a theory of matter not yet delivered. \cite[p. 180]{Lehmkuhl_genesis}
\end{quote} Here we seek only to invoke   a distinction between the natures of the gravitational and electromagnetic fields: a meaning closer to the folklore than that intended by Einstein.     } Nonetheless, in both theories, we can only objectively and unambiguously characterise the waves  in similarly special circumstances. Both the electromagnetic and the gravitational fields come to us as whole, and we must carve out  the  part that corresponds to radiation; but the joints are, in both cases, only apparent at very large distances.

As  to the energy carried by the waves, we find  a happy  overlap of regimes. 

First, it is important to state upfront that an energy-momentum tensor is well-defined for electromagnetism, but it is not, in general, for the gravitational field. Therefore we can always define the local energy of the electromagnetic field at a spacetime point and in a given frame, while in general we cannot do so for the gravitational field.\footnote{The energy momentum tensor is locally, covariantly conserved: this is a consequence of the definition of the energy-momentum tensor via a diffeomorphism invariant Lagrangian which is a sum of a part that depends on the matter content and a part that doesn't; conservation holds irrespectively of  the Einstein equations.  The corrresponding conservation laws for the gravitational field are mathematical identities, known as the Bianchi identities: these are satisfied automatically if the Riemann tensor is defined in the standard way as a function of the metric.  } 

  Moreover, our intuitions for electromagnetism are mostly based on its applications in  Minkowski spacetime, where the energy-momentum tensor of the electromagnetic field is not only well-defined but conserved.  In gravity, we can associate energy to the gravitational field only in certain regimes. For instance, we can do so in the weak-field regime of general relativity.  In this regime the analogy between gravity and electromagnetism is very strong: 
we can treat perturbations of the gravitational field in much the same way as we treat the electromagnetic field: as a kind of matter field on a rigid background geometric structure. For these gravitational perturbations, we can infer the same conclusions about radiation and energy conservation as we do for electromagnetism.

In a generically curved spacetime, as is well known, the notion of conserved energy is, to say the least, complicated. Generically,  energy conservation  is not locally meaningful---irrespective of whether that energy refers to gravitational or electromagnetic waves. If we allow geometries to vary arbitrarily in some region we can still make sense of the energy of the gravitational field by assuming spacetime to be  asymptotically flat. In that context,  we can interpret the spacetime as representing an isolated subsystem. The energy of the entire spacetime is interpreted as  the total energy of the subsystem in the fulness of time; and we can also interpret the flux of radiative energy `carried away' to infinity from the system as time passes. As Bondi, Penrose and others showed, gravitational waves objectively carry energy away to asymptotic null infinity; as \cite{ADM} showed, the entire energy of a spacetime is registered at asymptotic spatial infinity; and finally, as Geroch, Ashtekar and others showed,   the energy carried away to null infinity can be seen as part of the energy contained in the entire spacetime.\footnote{We are here avoiding the question of what  kind of data on a compact subset allows an asymptotically flat treatment.  For spatial infinity, there is considerable freedom in the initial data one can glue   to an asymptotically Kerr spacetime: see \cite{CarlottoSchoen2016} for a review. And thus the context in which ADM energy is well-defined is well-understood. But for asymptotic null infinity, the question involves teleological conditions on the full evolution of initial data, which is less understood. \label{ftnt:gluing}}

But asymptotic infinity is a hard place to get to. So why is it so useful in general relativity, even for experimental predictions? Because it is how we represent isolated subsystems in theories with long-range forces, like electromagnetism and gravity.  Fortunately,  in some circumstances, given a frame and appropriately separated relative scales, we can also identify gravitational energy and its transfer to bodies, even without going `all the way out' to infinity. For instance, the scale separations between the size of our LIGO detectors and the distance to what it is observing suffice for us to identify a component of the gravitational field here that represents the outgoing radiation emitted by the astrophysical sources. Or, in the sticky-bead example (see Section \ref{sec:bead}), the scale of the bar, the beads, and the curvature already allow us to distinguish a radiative component that does work:  we can verify that  the gravitational wave will have less energy at infinity than if it had not encountered the bar. 



In sum, {\em we} will present two main arguments in favour of the received view about the energy of gravitational energy. First, whether or not spacetime is asymptotically flat, radiative energy transfer---a redistribution of the total conserved energy into identifiably distinct components of a solution of the theory---occurs, or fails to occur, in conceptually similar circumstances for the gravitational field as for the electromagnetic field. Thus we will argue that sceptics who maintain that we don't understand energy transfer for gravitational radiation, must also claim we  don't understand it for electromagnetic radiation; even if the  \emph{energy-momentum tensor} is generically well-defined for  electromagnetism but not for gravity.  Second, in the case of gravitational radiation, because of the compounded subtleties of coordinate invariance and non-linear field equations, the question of carving out the objectively radiative components of the gravitational field was more explicitly tied to the question of whether these components carried energy. Heuristically, defining a wave requires  a rigid background and this same background can be used to define energy conservation. .

We thus conclude that though there are many  dissimilarities between gravitational radiation and electromagnetic radiation, they do not  license a relevant distinction with respect to  energy transfer.   The fact that the energy of the gravitational field is not \emph{generally} well-defined is consistent with both points of our conclusions: that  the notion of energy carried by a wave is valid in a (very relevant) regime of general relativity; and that gravity and electromagnetism swing together in this respect: the  effective notion of gravitaional radiation that emerges in this regime is as well-defined and as part of the basic furniture of the world as electromagnetic radiation.
 
 Here is how we plan to proceed. In Section \ref{sec:technical} we give a succinct list of technical results that we will invoke in this paper (we give a more detailed account in the appendix). In Sections \ref{sec:EM} and \ref{sec:conserv} we  address the recent philosophers' scepticism: in Section \ref{sec:EM} we will compare gravitational and electromagnetic radiation and in Section \ref{sec:conserv} we discuss energy conservation and isolated subsystems. 
 
 \section{Technical results}\label{sec:technical}
 Here we  remind the reader of the following crucial points about electromagnetism and gravitational waves ( labelled for  later reference):
 
 First, regarding the wave behavior of the fields:
 \begin{enumerate}[R(i)]
   \item Both the gravitational and the electromagnetic field obey constraints: these are equations that are not dynamical, but must be satisfied by any valid initial data for the theory. For instance, one of way of parsing the electromagnetic field according to constraints and dynamics is to say that the field has a component that is determined by the simultaneous distribution of charges and one component that has its own dynamics; loosely called `radiative'.  
   
   In  vacuum, in a contractible space, due to its linearity, the entire electromagnetic field can be unambiguously characterised as radiative. 
  But in general, e.g. when compact sources are present, there is no single local decomposition of either the gravitational or the electromagnetic fields into a radiative and a `Coulombic' part:  for a given region of a generic spacetime, different decompositions can lead to different conclusions about radiation. 
  
  \item The `waving part' of an electromagnetic field is  unambiguously identifiable  far away from the source, as the `component of the field' that falls-off as $1/r$;   the region in which this behavior occurs  is called \emph{the wave-zone}.

 In asymptotically flat spacetimes, we can use the Penrose-Newman null tetrad formalism to directly characterise the different components of the electromagnetic field that fall-off at the different rates: this is the electromagnetic Peeling Theorem. This theorem allows us to identify the components of the field that `become more and more' representative of electromagnetic radiation as one moves away from the source: these are the scalars $\Phi^2$.

 \item Similarly to the electromagnetic case, in the weak field limit of vacuum general relativity, in which we ignore everything but linear perturbations of the Minkowski metric, and  in a particular choice of gauge (transverse-traceless), the metric perturbations satisfy the usual wave equations on the Minkowski background and thus are entirely radiative. In this gauge, it is clear that there are two propagating degrees of freedom of the perturbation, as expected from the canonical counting of the physical degrees of freedom of the gravitational field. If the perturbations are sourced by matter, precisely as in the electromagnetic case, we can  identify the radiative parts of the perturbation as those components with the appropriate $1/r$ fall-off, in the wave-zone. These components depend on the quadrupole moment of the sources. 

 \item  Moving away from the weak field limit we have to deal with the non-linearities of the Einstein field equations head-on.  But there are no dimension-length constants in the theory that could characterize the onset of the strong field regime in general relativity.
  This implies the strong field regime is not associated with a particular length scale, and instead can be reached at any scale if some characteristic radius representing curvature becomes ``small". It is this comparison of length scales that  justifies the extrapolation of idealised asymptotic features of general solutions to finite distances in individual solutions. 
  
\item As in the electromagnetic case, in asymptotically flat spacetimes, we can use the Penrose-Newman null tetrad formalism to directly characterise the different components of the gravitational field that correspond to gravitational radiation. The Peeling theorem for the curvature also let's us identify components that fall off as $1/r$, and that, asymptotically, become `more and more' like the radiative modes of the weak-field approximation: these are the Weyl scalars $\Psi_4$. In the bulk of the spacetime, the scalars  $\Psi_4$ strongly depend on a choice of null tetrad basis. In certain algebraically special spacetimes these choices can be physically constrained and the $\Psi_4$ can be taken to correspond locally to gravitational radiation in a strict sense. But generically, an unambiguous notion of radiation is only available at asymptotic infinity (see \cite[Ch. 6-8]{Dambrosio_grav}).\footnote{The original definition that encapsulates the modern usage can be traced back to \cite[p. 333]{BondiSachs_gravIV} ``A covariant characterization of spaces that are free of mixed radiation at light-cone
infinity has been suggested. The characterization seems to agree with every
thing
that is known about gravitational radiation fields at present. The final definition
proposed [is] that a field with asymptotically geodesic rays is one that is free of
mixed radiation at large distances...''} 
 
 \end{enumerate}

Next, regarding the conservation of energy:

\begin{enumerate}[E(i)]

 \item Given the background Minkowski metric and its associated Killing vector fields, one has meaningful notions of energy conservation for matter fields (i.e. the right-hand-side of the Einstein field equations), which  apply equally  to the electromagnetic field and to the   linearised gravitational degrees of freedom. More generally, conservation laws can be deduced for spacetimes that are suitably algebraically special. But for a generically curved spacetime, no covariant, quasi-local conservation laws exist without the introduction of some background structure.\footnote{The canonical reference here is \cite{Szabados2004}, who states (p. 9): ``contrary to the high expectations of the eighties, finding an appropriate quasi-local notion of energy-momentum has proven to be surprisingly difficult. Nowadays, the state of the art is typically postmodern: Although there are several promising and useful suggestions, we have not only no ultimate, generally accepted expression for the energy-momentum and especially for the angular momentum, but there is no consensus in the relativity community even on general questions  (for example, what should we mean e.g. by energy-momentum: Only a general expression containing arbitrary functions, or rather a definite one free of any ambiguities, even of additive constants), or on the list of the criteria of reasonableness of such expressions.'' } 
 
 \item Energy transfer is not solely radiative. In the case of electromagnetism in a Minkowski spacetime, energy transfer through a surface is given by the flux of the Poynting vector. But the Poynting vector can be non-zero for a non-radiative source: a Lorentz boost of a purely Coulombic field gives rise to both an electric and a magnetic field and to a non-zero Poyinting vector; but that field is not radiative. In accord with item R(ii),  the magnitude of such a non-radiative Poynting vector falls-off with distance faster than a radiative Poyinting vector; indeed,  the former's flux vanishes at asymptotic infinity, unlike the latter's.
 
 \item In the asymptotically flat case, using the Penrose-Newman null tetrad formalism, we can again identify different fluxes at infinity. We have conserved charges at asymptotic spatial infinity that correspond to the integral of  non-radiative components. For instance, for electromagnetism and gravity we get the total electric charge and the ADM energy-momentum, respectively. And we have an energy flux that corresponds to an integral over null asymptotic infinity whose arguments include only radiative components. One can interpret the difference between the ADM energy and the energy of the radiation up to a certain (retarded) time as the `leftover' energy of that spacetime at that time \cite{AshtekarMagnon_energy}.
 
 \end{enumerate}

\section{ Gravitational radiation in thought and in reality}\label{sec:EM}

We will now deploy Section \ref{sec:EM}'s similarities between gravitational and electromagnetic radiation in discussing two iconic, and historically significant, ``experiments''. One is a thought-experiment (Section \ref{sec:bead}); the other is a decades-long observational programme (Section \ref{sec:pulsar}). (For historical details, cf. again e.g. \cite{Kennefick2007}). In the course of this, our disagreements with the sceptics about radiative gravitational energy will become clear, since they also focus on these two cases. 

\subsection{The sticky bead}\label{sec:bead}
After disavowing his theoretical discovery of gravitational waves in the 1930's, Einstein was set right  by Infeld and (indirectly) Robertson. With their help, he became convinced  that he had misinterpreted the coordinate artefacts  of his and Rosen's construction. Rosen, his collaborator on the original disavowal, was less convinced, alongside many others. 

The state of play remained relatively inconclusive until the famous 1957 conference in Chapel Hill, in which Feynman gave his famous ``Sticky Bead'' thought-experiment. Leading up to the conference,  Pirani and Robinson had been emphasising the role of tidal effects and of the Riemann curvature in providing a coordinate-independent description of gravitational radiation. Using these ideas, Feynman pictured a rigid rod with two ring-like beads, free to
slide with friction on the rod, placed in the path of a gravitational wave. Thus,  since the beads would slide back
and forth on the bar, and through the action of friction heat up the bar, Feynman concluded that
the bar can only heat up if the gravitational waves transfer energy to it. \citet[p. 197]{Rovelli_woods} expresses the idea forcefully:
\begin{quote}
A strong burst of gravitational waves could come from the sky and knock down the rock
of Gibraltar, precisely as a strong burst of electromagnetic radiation could.\footnote{The quote continues: ``Why is the first
“matter” and the second “space”? Why should we regard the second burst as ontologically
different from the second? Clearly the distinction can now be seen as ill-founded.'' Here  we wish to remain agnostic about this stronger statement. We can admit a significant ontological difference between spacetime curvature and other forces---e.g. electromagnetic---while still taking radiation in each theory to be equally capable of trasmitting energy.}
\end{quote}
But  many philosophers demur. For instance, \cite[p. 30]{Duerr_grav} writes:
\begin{quote}
Therefore, even if [...] we did register an
increase in thermal energy of a Sticky Bead detector, we wouldn't
be licenced to infer a transfer of energy from the GW, so as to restore
energy balance. Rather, it would seem more natural to accept an
alternative stance: Energy conservation simply ceases to hold in GR.
The detector would just heat up - without there being a causal story
about it that would allow us to track the lost energy. Energy conservation
is just violated (quantifiably!), when a GW hits a detector.
\end{quote}
 Other arguments, in recent years common among philosophers, proceed in a similar spirit. As is often pointed out, one can realize the relevant sliding motion of the beads through any geodesic
deviation; e.g. in the exterior Schwarzschild metric, as the beads fall towards the center of the planet.

 Before criticising these arguments, let us first make a concession.  It is true that, had we placed the beads close to the source, it would be impossible to univocally distinguish the energy they obtained solely from gravitational radiation, since, by item R(i) (see also R(iii)), at that distance, there is no unique, unambiguous decomposition of the curvature into an outgoing radiative and a non-radiative component. This is analogous to placing an electromagnetic antenna very close to an oscillator, in the near-field zone (see R(ii)): there is definitely a changing electromagnetic field that does work on the antenna, but what part of that work is solely due to radiation?\footnote{ Indeed, as the excellent Wikipedia entry on Electromagnetic Radiation correctly states: 
\begin{quote}the term ``radiation" applies only to the parts of the electromagnetic field that radiate into infinite space and decrease in intensity by an inverse-square law of power, so that the total radiation energy that crosses through an imaginary spherical surface is the same, no matter how far away from the antenna the spherical surface is drawn. Electromagnetic radiation thus includes the far field part of the electromagnetic field around a transmitter. A part of the ``near-field" close to the transmitter, forms part of the changing electromagnetic field, but does not count as electromagnetic radiation.
\end{quote}}

In the same manner, we believe all parties would agree that   talk about energy transfer from gravity to a  sticky bead, or to a glass of water,   only requires a regime where the sticky bead  or the glass of water itself can be well approximated in standard non relativistic terms (in their frame).   Then they have a well defined local energy, which is measurable, and if this energy grows and their only interaction is gravitational, it is legitimate to say that there was transfer of energy from gravity to the object.

But the question is whether that transfer is due to an objectively defined gravitational wave, and whether, when it is, if the approximations required   also allow us to attribute energy to that wave.

We contend that  the  assumption underlying Feynman's picture of a  discernible  gravitational wave hitting the sticky bead is equivalent to an assumption of the weak field regime or of sufficient distance from the source. That is, the beads are at a distance such that, according to R(iv), the Newman-Penrose components $\Psi_4$---which are defined everywhere but retain some frame dependence that is  irrelevant under the intended interpretation of asymptotic infinity---provide good approximate notions of outgoing radiation (as in \citep{BondiSachs_gravIV}; see \cite[Ch. 8]{Dambrosio_grav} for a pedagogical review). These notions are approximate in the sense that the difference between their values where the beads are  and their asymptotic values is smaller than some quantity---e.g. the experimental error bars---and this difference decreases monotonically with distance. In other words,   for each particular spacetime model of the scenario, we can compute $\Psi_4$ at ever farther geodesic distances from the source, on a frame that has a well-defined physical interpretation, and verify that its difference to the value taken asymptotically is bounded by some relevant (e.g. experimental) limit. 

Of course, spacetime could fail to admit the relevant notion of ``increasing distances'', or otherwise fail to admit discernible, frame-independent gravitational waves. The  possibilities for spacetime geometries are enormous after all, and have little respect for the geometric intuitions we get from our tame surroundings. It would be absurd to require that generic spacetimes should admit clearly discernible  gravitational waves; and yet it is virtually guaranteed that extended bodies placed in those spacetimes would be subject to changing tidal effects, and thus to tension and work.  As for the beads, they could heat up the bar via tidal forces related to any arbitrarily varying (in their frame) Riemann tensor; this is guaranteed by the geodesic deviation equation \eqref{eq:geod_dev}, and no more.\footnote{We could even split the Riemann tensor into its Ricci and its Weyl parts: the latter is taken as independent of the matter content, and thus usually associated to the purely geometric degrees of freedom. }  Absent  any background structure, including asymptotic conditions, we would not know how to define the spacetime's energy (as per item E(i)).  In these cases we can agree that the beads' gain or loss of energy---however it is defined in the beads' frame---are due to tidal forces, even if it is meaningless to say that the energy encoded in the spacetime curvature has diminished or increased. 

Nonetheless,  in either of the two  regimes where \emph{we can} clearly discern outgoing radiation, Duerr is mistaken to say that we cannot tell a causal story about transfer of energy from the  wave to the detector. In these regimes, the accounting of energy is explicit, just as it is with electromagnetic radiation. Thus a gravitational wave will have less energy at infinity than if it had not encountered the bar. The point here is that if one assumes outgoing gravitational radiation has been emitted from a body and can be discerned,  one is bound to give an account of that radiation. And the cases in which we know how to do that  either require algebraically special spacetimes, with associated conservation laws; or  asymptotically flat spacetimes, where the phenomenon takes place very far from the source, where asymptotic conservation laws are approximately observed.  

 Turning to the second argument: it is true that beads that are freely falling in a Schwarszchild background  \emph{would} slide---not necessarily back and forth, but creating friction nonetheless. That is, as described above, the gravitational field would impart energy to the bar even in a stationary spacetime, such as Schwarzschild. Since it is agreed by all that such vacuum spacetimes do not carry gravitational waves, the sticky bead argument, \emph{by itself} gives us no reason to believe that it is the gravitational waves that are transferring energy.

So we submit that  this appraisal gets things backwards.  The sticky bead argument never claims that radiation is the sole purveyor of gravitational energy transfer:  as described in item E(ii), the fact that energy transfer occurs non-radiatively is no mystery, and it is equally true in the case of electromagnetism, where we can obtain a non-zero Poynting flux from electro-magneto-stationary sources.  If we assume \emph{no} gravitational wave is present, obviously its causal powers must also be absent. But here we are  assuming the existence of a discernible gravitational wave to begin with, and that it is the only source of tidal effects. 

In sum, so as long as we picture the sticky bead in the weak field approximation of general relativity---as it was meant---gravitational energy transfer \emph{is} solely radiative. Leaving the weak-field approximation, the same conclusion holds to ever higher degrees of approximation as we place the sticky beads farther and farther from the source. As described above, given the relative scales, we need not even involve the idealisation of asymptotic infinity directly (see item R(iv)).

\subsection{The binary pulsar}\label{sec:pulsar}


In 1974, almost two decades after the introduction of the sticky bead argument,  Russell Hulse and Joseph Taylor discovered a
binary star system (now known as the Hulse-Taylor binary) consisting of a neutron star and a pulsar, which emitted regular pulses detectable on Earth. Theoretically, that system would be a source of gravitational waves. The question then was, would the quadrupole formula give a reasonable approximation of the source strength of this system? Between then and the early 1980's, joint efforts by many theoreticians---most notably Clifford Will and Thibault Damour, who introduced a new method especially
tailored for computing the third-post-Minkowskian,
gravitational field outside two compact bodies---culiminated in a successful calculation, making precise predictions in a post-Newtonian approximation up to fifth order ($c^{-5}$). The precise mathematical results  about the orbital decay agreed with exquisite precision both with the standard,  quadrupole
formula for gravitational radiation in the weak field limit and with the observations of the Hulse-Taylor pulsar. The evidence pointed directly to radiative energy transfer.\footnote{But these computations do not give precise predictions for the gravitational wave-forms: for that, we need numerical methods in full general relativity. This is a distinct and enormously complicated task, whose breakthrough moment came much later, in  \cite{Pretorius_2005}.  }

Still,   in recent conversations and in print, some philosophers demur. 
One argument, the more easily countered, is that: 
\begin{quote} the pulsars (modelled as dust particles) are
in free-fall. Hence they move inertially. Shouldn't their kinematic
state therefore remain unaltered?   \citet[p. 26]{Duerr_grav}\end{quote}
 But the more common and (superficially) convincing argument against the received wisdom about the binary pulsar is that we can bypass explanations employing energy transfer by resorting to numerical simulations for solutions of the Einstein field equations directly.

Let us take these comments in turn. For a freely-falling cloud of pressureless dust particles, one can find an adapted coordinate system in which the position of the particles don't change. Without any calculation   \cite[Sec. 2.1]{Duerr_grav}, we can conclude that the naive notion of kinetic energy adapted to this coordinate system cannot change. 

First, we point out that under the motion induced by a passing gravitational wave what changes are relative positions of particles, not their individual velocities. This relative motion is best described by the geodesic deviation equation. For $v^a$ the tangent vector to the time-like geodesics, and $r^a$ the transversal displacement vector, we get, an entirely covariant, non-zero acceleration: 
  \be\label{eq:geod_dev}
  \frac{\D^2 r^a}{\d t^2}:=v^c\nabla_c(v^d\nabla_d r^a)=R^a_{\phantom{d}bcd} v^b v^d r^c.
\ee
So under a frame that is adapted to the time-like geodesics, we can easily associate a non-trivial notion of kinetic energy to the accelerated  \emph{relative} motion.

Second, the fact that motion is geodesic requires further analysis in order to conclude that it emits or absorbs, or fails to emit or absorb, radiation, or whether that provides a suitable explanation.\footnote{Here is a proof of principle for doubting a necessary connection: in the Kaluza-Klein framework electromagnetic forces are  geometrised. In that framework, charged particles undergoing motion in a background electromagnetic field are interpreted as following geodesics, and nonetheless absorb radiation.}
 
Indeed, as  mentioned in the introduction, Rohrlich and Fulton  showed in  the 1960's that  a freely falling charge could be taken to radiate, in Earth's frame, but it would have no detectable radiation to a comoving observer.  But the matter here is subtle, and involves global properties of the accelerated observer in Earth's frame and of the inertial charge.\footnote{The main difficulty in locally analyzing the radiation emitted by an inertial charge in the context Rohrlich and Fulton discussed is the fact that, in general, because the current associated with such a charge does not have a compact support, it cannot be completely confined in any Rindler wedge. A different definition of inertial charges---that takes the infinite limit of acceleration and thus confines these sources to the Rindler wedge---concludes that these accelerated observers would find \emph{no} radiation.} 
 Thus, in their review article,  \cite[p. 2]{Saa_electron} write: 
\begin{quote} We need to recognize that
the concept of radiation has no absolute meaning and depends
both on the radiation field and the state of motion of
the observer.\end{quote}


 The second argument, about the explanatory usefulness of radiation, can be illuminated by  the question of ``radiation-reaction'',  originally discussed by Dirac and DeWitt and Brehme. 
 Let us elaborate. 
 
  A charged particle or an extended or spinning body---like the components of the binary pulsar---can't be taken to follow the paths of  neutral point-like particles, namely,  the geodesics of a background spacetime. For unlike neutral point-like particles, these objects necessarily contribute to the energy-momentum tensor, and thus change the spacetime geometry.  The approximation schemes used to find their true trajectory given a background geometry are precisely what we mean here by \emph{radiation reaction}. The idea is that the background geometry scatters the electromagnetic, or the gravitational field, sourced by the particle's whose motion we are trying to determine.  In more detail, an electromagnetic field originates on the charge in the past, is scattered by a gravitational field some distance away, which then  produces a non-zero force in the present. Though the equation determining the deviation from the comparative motion of the uncharged particle will involve non-local contributions--- through the value of Green's functions acting on the (retarded) past of the particle---there is nothing mysterious in the non-local character of the force. It is the result of reducing the interaction between fields (gravitational and electromagnetic) to a finite dimensional description in terms of the source's motion alone  (see \cite{QuinnWald} and \cite{Poisson2011} for a comprehensive review).
  
 And again, in certain  contexts in which we can define suitable local conservation laws, namely, in a globally hyperbolic, stationary spacetime, we \emph{can} use the radiation reaction to give a precise account of quasi-local conservation laws.\footnote{In the gravitational case, one can only explicitly show this for the linearised theory around Minkowski spacetime, but in principle the method could be extended to include the linearised theory around the Kerr family of solutions \cite[p. 23]{QuinnWald}.} As described in \cite[p.3]{QuinnWald}: ``This provides justification for the use of energy and angular momentum conservation to compute the decay of orbits due to radiation reaction.'' Reference to radiation and scattering is crucial to this explanation. How could we deligitimize the use of such effective terms within a theory without condemning the vast majority of physical concepts to the same fate?

 We will come back to this topic about the concepts of physics in the Conclusions. But for the main message of this paper we don't need to be so general: the explanatory utility of the concept of radiation is again analogous for the electromagnetic case; the strength of the analogy---and the ubiquituous use of electromagnetic radiation in physics---suffices to shift the onus of explanation to the sceptic about gravitational radiation and its energy transfer. 
 
  Being explicit: Yes, it is true that the motion of the LIGO detector plates can in principle be described without ever using the notion of  energy transfer.  Such motion could  be entirely accounted for by free fall and violations of free fall due to non gravitational forces.   But the same formal manouvers are available in electromagnetism. So, if we discard an ``explanation" because there is in principle an account that is more general, then we must discard energy transfer as an explanation equally for gravity and for electromagnetism.  Conversely, if we count  the use of  regularities that hold in special regimes as explanatory, then energy transfer by waves is equally explanatory in general relativity and in electromagnetism.

\section{Asymptotic flatness and dynamical isolation}\label{sec:conserv}

Lastly, we turn to the role of asymptotic flatness in considerations of energy conservation. This is a last resort for the sceptic, who may try to bite the bullet and deflate the ontic significance of any wave's energy transfer on generic curved spacetimes.
\cite{Duerr_grav} writes: \begin{quote}
[p. 30] Asymptotic flatness would have to be shown to be a
“working posit” of (i.e. essential for) relativistic astrophysics. But
this is questionable. ...

[p. 34... asymptotic flatness is]  an idealisation in
Norton's sense: The embedding spacetime is an unrealistic, surrogate
spacetime. Consequently, realism about notions of gravitational
energy based on asymptotic flatness isn't straightforward. \end{quote}

 The argument here is that, since asymptotic flatness is not fundamental  in general relativity, any concept or quantity that depends on this assumption must also be less than fundamental in the theory. This is a blunt argument, condemning  our understanding of  energy transfer, \emph{simpliciter}, in generic curved backgrounds.

 And, amongst philosophers, we have witnessed a distinct, frequent  misunderstanding of the meaning of the asymptotic integrals that are involved in the definitions of the relevant energies.  Namely, that different notions of energy are `holistic', that they cannot distinguish between the energies due to gravitational waves and due to other, non-radiative contributions. For instance, in a recent talk \cite{Fletcher_talk} describes the situation thus:\footnote{ We do not mean to single out Fletcher for what may not be his considered views. But this passage illustrates concisely what seems to us a widespread view among philosophers of physics. }
 \begin{quote}
It is extremely tempting, on the story that I have given, to say that because we find that the Bondi energy decreases with time, that gravitational radiation carries away positive energy from a radiating system. But [...] we should resist getting carried away,  because strictly speaking gravitational waves don't have any Bondi energy of their own. [...] 
these global notions of energy are assigned to whole space times, so we can't divide the energy content into one part which is associated with one part of a space-time and another [...] 
with [...] the gravitational waves.   [...]   therefore we can't say that the gravitational waves have Bondi energy that is carried away. All we can say is that the Bondi energy decreases.\end{quote}

Are they right? Let us once again take their claims in turn: first Duerr and then Fletcher.

Is asymptotic flatness a working posit for astrophysics?  Agreed: not in full generality, but it surely is a working posit to study gravitationally isolated subsystems. What undergirds the assumption of asymptotic flatness is just dynamical isolation of subsystems; conceptually, dynamical isolation is what grounds both an unambiguous separation of radiative and Coulombic modes and conservation of energy. To talk about energy transfer, we need to be able to clearly distinguish subsystems within the theory. And again, this condition (of dynamical isolation) is necessary to discuss energy conservation \emph{in general}---even in  the familiar case of Newtonian mechanics---not just in general relativity.  

In general relativity we cannot set the gravitational field to zero at a supposed boundary between subsystem and environment. What we can do is demand  that gravitational (tidal) forces become less and less pronounced at far enough distances from the subsystem. Once again, there is no decree in the theory that every spacetime should have  subsystems that are sufficiently isolated: the very idea of removing oneself farther and farther from a subsystem can fail. But if we want to talk about concepts that require dynamical isolation, such as radiation and energy, we have no other recourse. Here is \citet[p. 182]{Penrose_gr_problems}  making this exact point: 
\begin{quote}
``Asymptotically flat spacetimes are interesting, not because they are thought to be realistic models for the entire universe, but because they describe the gravitational fields of isolated systems, and because it is only with asymptotic flatness that general relativity begins to relate in a clear way to many of the important aspects of the rest of physics, such as energy, momentum, radiation [...]
\end{quote}
 
 Indeed, this kind of assumption extends even to Newtonian mechanics.  There, to apply the laws and obtain conservation of energy, we must describe the system in an inertial reference frame. But how do we ensure our description is  in an inertial frame?

 Corollary IV in Newton's Principia says that, \emph{if we can ignore} external influences on some subsystem, the center of mass of said subsystem will move uniformly with respect to absolute space (or be at rest).\footnote{“The common centre of gravity of two or more bodies does not alter its state of motion or rest by the actions of the bodies among themselves; and therefore the common centre of gravity of all bodies acting upon each other (excluding external actions and impediments) is either at rest, or moves uniformly in a right line..”} Jointly with Corollary V,\footnote{``The motions of bodies included in a given space are the same among themselves, whether that
space is at rest, or moves uniformly forward in a right line without any circular motion.''
} we conclude that, \emph{if we can ignore external influences} and are not in circular motion,  we can for all practical purposes treat the center of mass of our subsystem as being at rest with respect to absolute space. The underlying assumption of `dynamical isolation' here is that our subsystem is sufficiently far removed from other, external, bodies. Is this a fully general presupposition of Newtonian mechanics? Once obtained, will it obtain for all time? `No' is a conceivable answer to both questions. Nonetheless, the presupposition is necessary for conservation of energy, along most other practical applications of the theory.\footnote{`Ignoring external influences' is  subtle business: it does not necessarily mean that all external forces on a subsystem have to vanish.  As argued in \cite{saunders2013}, to empirically apply the laws,  Newton has to implicitly resort to Cor. VI, which says that: “If bodies, any how moved among themselves, are urged in the direction of parallel lines by equal accelerative forces; they will all continue to move among themselves, after the same manner as if they had been urged by no such forces.''. So we can empirically apply the laws if our subsystem is sufficiently distant from external sources so that they would act equally (in parallel) on all of its components.}

 Let us now turn to the second type of misunderstanding, about the holistic nature of asymptotic notions of energy.  A common mistake is to take the relevant notions of energy to arise from integrals over entire Cauchy surfaces---for ADM energy-momentum---or even `from slices that don't intersect radiation escaping out to infinity'.  In truth, the relevant quantities are strictly integrals over either asymptotic null infinity or over asymptotic spatial infinity.\footnote{Of course, one could use Stokes theorem to convert these integrals into an integral of different dimension, but the details of the fields on the further dimensions would be irrelevant to the value of the integral.} And so the relevant integrals  are already calculated over a surface where it is possible to uniquely distinguish the radiative from the non-radiative components. 

 The integrated Bondi  energy flux is given in \eqref{eq:GWFluxes} in Appendix \ref{app:grav}; the ADM energy is better-known, but would require the introduction of terminology that is besides the point of this paper. 
 The ADM energy  is usually interpreted as the total energy available in the spacetime. As it is a quantity calculated at spatial asymptotic infinity, it is `static', and does not evolve, unlike the Bondi energy. So what is the relation between the Bondi and ADM energies? Consistently, the Bondi energy can be interpreted as the energy remaining in the spacetime at the ``retarded  time" after the emission of gravitational radiation.  For, as shown by \cite{AshtekarMagnon_energy}, the Bondi energy at a certain cross-section of $\mathscr{I}$ differs from the ADM energy by the integral  \eqref{eq:GWFluxes}, up to the retarded time given by that cross-section. So Fletcher may be right that the Bondi energy does not distinguish the energy due to radiation, but it is the \emph{Bondi-energy flux} that can be seen as `subtracting energy' from the spacetime. Thus the difference of Bondi energies at two different times is unambiguously associated to the energy of the radiation that leaves spacetime in that interval.

 In sum, the radiated energy depends only on the components that encode gravitational waves, as understood both in the linearized and asymptotic limit.  And these different notions of energy are remarkably consistent: if we understand the energy at spatial infinity as the energy of the entire spacetime, we can understand a difference between the total energy at spatial infinity and the energy radiated away along null infinity up to a given retarded time as  the  energy left in the spacetime at that given (retarded) time. Thus, if a part of the gravitational wave is absorbed and turned into e.g. thermal energy, we will find a corresponding subtraction in the energy radiated away to infinity. 
 
  Again, an analogy applies to electromagnetism:  the Gauss law gives us the total charge in the spacetime; this is an integral over spatial infinity. A different  integral over different components gives us the radiated electromagnetic  energy from the spacetime.\footnote{The analogy here is only limited by the fact that the total charge is constant, since it  does not track a loss of energy and there is no flux of charges at infinity.}

 Thus, contrary  to a straightforward interpretation of Fletcher's passages, asymptotically, we \emph{can precisely} separate the energy of the system  into one part which is associated with the gravitational waves and one part that is related to other charges of the isolated subsystem.
 
  And while it is true that constants such as the ADM mass of a spacetime may  not discern e.g. whether a given spherically symmetric solution has singular behavior (i.e. vacuum Schwarzschild) or a star at its centre, there is no reason to think this is problematic for any of the concepts discussed here.\footnote{Indeed, as \cite{MisnerWheeler1957} have long ago argued, with sufficiently complicated topologies, we can even trap electric fields in wormholes and thereby obtain `charge without charge'. }
    
   \subsection{Why can't we define gravitational waves in the bulk of a generic spacetime?}

Finally, let us address a possible point of confusion: why couldn't we take  $\Psi_4$, described in Appendix \ref{app:grav},  to describe  gravitational radiation,  for any region of a generic spacetime? The short answer is that the choice of null-tetrad is arbitrary, and different choices can change the values of the Weyl scalars. But in certain spacetimes, there are physically significant choices, that can uniquely determine the values of (some of) the Weyl scalars. 

This is the case of  algebraically special spacetimes that characterise  gravitational radiation spacetimes as being of Type N. In this case, there is a particular choice of null direction $k$, representing the direction of the wave, such that \be\label{eq:Weyl_n} C_{abc}^{\phantom{abc}d}\,k_d\,{=}\,0.\ee
 For these types of spacetimes, we find, \emph{for some compatible choice of null tetrad}:
 $$\Psi^0=\Psi^1=\Psi^2=\Psi^3=0.$$
  See \cite{Szekeres1965} for a derivation of the explicit geometric relation between $\Psi_4$ and gravitational radiation in Type N spacetimes (and for the geometric interpretations of the other Weyl scalars discussed in Appendix \ref{app:grav} as well).
  
  One  important point for this paper is that, since these solutions  are algebraically special, they will come with some background structure which can be used to define conservation laws (see e.g. \cite{Aksteiner_2021} for a recent review of conserved quantities in algebraically special spacetimes).\footnote{However, we should point out that, unlike the case of gravitational waves in a general spacetime, the discovery of these special waves was \emph{not} linked to conservation of energy.}

Going back to the generic case of asymptotically Minkowski spacetime,  using the smooth limit of the Schouten tensor to $\mathscr{I}$ and its relation to the Weyl tensor, we straightforwardly obtain, in that limit, a constraint equation of precisely the form  \eqref{eq:Weyl_n}, saying that asymptotically, we approach a spacetime that (can) include a gravitational wave as understood in the algebraically special case.\footnote{In more detail, the entire physical Weyl tensor vanishes at $\mathscr{I}$. What one uses instead for these computations is the asymptotic Weyl tensor, defined as $K_{abcd}=\Omega^{-1}C_{abcd}$. Since $\ell_a=\Omega^2\, \hat\ell_a$, we still have $K_{abc}^{\phantom{abc}d}\ell_d\rightarrow 0$.  }

Consistently, as remarked after equation \eqref{eq:Psi4inStrain},  the entire, coordinate independent conformal geometry of $\mathscr{I}$ is encoded in the conformally invariant limit of the shear, i.e. in the conformally invariant limit of $\nabla_a \ell_b$ as it approaches $\mathscr{I}^+$. Since it is encoded by a varying shear, radiation acquires a geometric---coordinate-independent---gloss.  In sum,  the   conformal geometry of $\mathscr{I}^+$ is entirely  determined by the radiative degrees of freedom; and conversely, by construing radiation geometrically we limit the role that different  choices of frame can have on its value.
\footnote{As shown by \cite{Geroch1977}, the (retarded) time derivative of the shear---the News tensor---vanishes in an asymptotically flat spacetime if and only if it is stationary; hence the name `News' is well deserved. Asymptotically, the first derivative of the shear  along $n$ (i.e. along retarded time) gives the News tensor, and its second derivative gives $\Psi_4$, which thereby inherits an invariant meaning, conditional on an intended interpretation of the directions represented by  $\ell$ and $n$ (see Equation \eqref{eq:Psi_shear}). } 

The moral is that asymptotic infinity gives us enough structure to unambiguously define  gravitational waves  because it is associated with a ``direction'' that is infinitely far away from compact sources: $n$ rules $\mathscr{I}$---it gives a choice of retarded times at $\mathscr{I}$---and $\ell$ is the radial direction away from $\mathscr{I}$ towards the bulk.  In the bulk, even if we can physically characterize $\ell$ and $n$, we would not be able to characterise radiation independently of the remaining choices of the frame/coordinate system.\footnote{There are many different ways to define `surfaces of constant retarded time' along $\mathscr{I}$. In Minkowski spacetime, one could construct a preferred one by sending null rays to infinity along a given time-like geodesic. But in a general curved spacetime, one could have several physical effects e.g. lensing, that make the ruling of $\mathscr{I}$ less operationally definite, which is why one usually goes the opposite way: defining a null-tetrad at $\mathscr{I}$ and propagating it inward. Supertranslations are part of the symmetry group of $\mathscr{I}$ and their effect can be seen as changing the surfaces of constant retarded time in a way that can depend on the angles of $\mathbb{S}^2$. Fortunately, by associating the relevant notions of radiation to the geometry of $\mathscr{I}$, i.e. to the trace-free part of the Schouten tensor at $\mathscr{I}$  (i.e. the News tensor in a particular choice of gauge---Bondi gauge), or to the limit of the shear, it is shown that the existence of radiation is independent of these choices. }

 
 

 \section{Conclusions}
 
 Even the philosophers that are sceptical about gravitational radiation  will concede that a gravitational wave is  special in that it has a long range: how could they not? But what these sceptics fail to realise is that this is a \emph{constitutive} property of radiation, and that it is shared by electromagnetic waves. Generically, both types of fields come to us as a whole; and at close range there are no apparent joints to be carved.  
  Nonetheless, there are conditions under which the joints become apparent---at great distances from the source for example---and these are the conditions under which we understand \emph{both} radiation and energy transfer. 
  
  In this paper, we have  not touched on topics in the philosophy of science that may be relevant to this issue. But one seems unavoidable, and indeed was briefly touched on at the end of Section \ref{sec:pulsar}. That is the idea that the only notions that are well defined are those that can be defined in all regimes encompassed by a theory; and that only such notions can claim to be part of the  furniture of the world (according to the theory). As we have emphasised,  the "local energy of gravity" doesn't fit that bill, since it is not defined in all regimes. Thus,  the sceptic will say, ``the local energy of gravity does not exist".  But physics rarely trades in fundamental ontology: it wouldn't have gotten much done  if it did! Physics, and indeed science more generally, trades in effective notions---like the energy of the wave, the horizon, the orbit, the black hole---that are perfectly well-defined for specific solutions, or for specific regimes, and that have exceptional explanatory value. A philosophy of science that denies any ontological status to  these notions  leads to an  impoverished picture of science, that we must reject. 
 
 In the case of general relativity,  gravitational waves had particularly turbulent origins, and were only accepted with the introduction of a fully invariant account. This account either required algebraically special spacetimes and perturbations therein, or  asymptotic infinity. In both cases, these further structures  could be related to notions of conservation, which further established the reality of the waves.  This is reassuring, and another example of the great unity of theoretical physics. But it would be a mistake to elevate this reassurance to a form of sanctioning:  our theoretical commitment to robust wave patterns is not \emph{conditional} on their being subject to energy conservation. Nor should we seek such sanction: `energeticism'---the XIXth century hope of reducing all natural phenomena to 'manifestations of energy'---is long dead in theoretical physics, and for very good reasons.  

   We finish with two quotes: both confirm our main message, even if the second aimed to sum up the very scepticism we are here rebuffing. 
   
      \cite[p. 100]{Dambrosio_grav} write:
    \begin{quote}
    The fluxes [see Eq. \eqref{eq:GWFluxes}] represent a landmark in the discussion on the existence of gravitational waves, which culminated in the nineteen-sixties. Since the inception of gravitational waves in 1916 by Einstein, there has been much debate about whether they are a real physical phenomenon, or whether they are a mere coordinate artifact. Eventually, this dispute was settled by the mathematical rigorous framework presented here, as it provides a \textit{gauge-invariant} description of gravitational waves. In particular, it provides a gauge-invariant description of the flux of energy and momentum carried by gravitational waves.\end{quote}

 And towards the end of his paper, \cite[p. 35]{Duerr_grav} writes: 
 \begin{quote} Three considerations bear upon the choice between failure of
energy conservation and energy transfer: 1. the contingency of
energy conservation on symmetries, 2. the existence of a satisfactory
formal account/representation of the energy transport, and 3.
the explanatory value of postulating energy transport rather than
energy decrease simpliciter, respectively.
\end{quote}
We have here shown that gravitational radiation does as well as electromagnetic radiation on all accounts.

 \subsection*{Acknowledgements} We would like to thank Jeremy Butterfield for valuable suggestions; HG would like to thank Harvey Brown and Simon Saunders for encouragement and suggestions; and Oliver Pooley, Patrick Duerr, and Fabio D'Ambrosio for discussions and feedback.

 \appendix 
 
 \section*{APPENDIX}
 
  This Appendix is based on \cite{Dambrosio_grav}. HG is thankful to Sebastian Muguertio Ramirez for pointing me to that source for this topic.
 
 \section{Electromagnetic radiation}\label{app:EM}
 Wave equations for the electromagnetic field are easily derived from Maxwell's equations, and so are  formal solutions to these equations in the presence of sources. Generally, the radiation field  should have three characteristic properties: it should oscillate, it should be transversal, and it should decay as $\frac{1}{r}$ as we move away from the source.

The difficult question is whether we can determine, locally, whether a given source, $J^\mu := (\rho, \vec{j})$,  generates a \textit{radiation} field. 
And the answer to the difficult question is that we can only determine whether a given source generates radiation asymptotically far away from the source.

Consider an electromagnetic source $J^\mu$  confined to a finite spatial region at the scale $d$. Given this source,  the vector potential that satisfies the Maxwell equations is: 
\begin{equation}\label{eq:formal_solution}
	A^\mu(t, \vec{x}) = \frac{1}{4\pi} \int_\Omega\d^3 x'\int_R\d t'\,\frac{J^\mu(t', \vec{x}')}{\|\vec{x}-\vec{x}'\|}\,\delta\left(t'-t+\|\vec{x}-\vec{x}'\|\right).
\end{equation}

The difficulty mentioned above arises from the fact that the source may have static parts which only produce Coulombic fields  and it may have radiating contributions. But the fields come to us as a whole: we do not yet know how to disentangle the different contributions. 

Thus, assume there is radiation and that it has a wavelength $\lambda = \frac{2\pi}{\omega}$. Moreover, assume an observer is located at the radial distance $r$ from the source (we will compare $r$ with $d$ later). 

We can decompose our source into static contributions, $J^\mu_\textsf{stat}(\vec{x})$, and radiative contributions, $J^\mu_\textsf{rad}(t, \vec{x})$, where the  radiative contributions are assumed, without loss of generality, to oscillate like ${e}^{-i \omega t} J^\mu_0(\vec{x})$. That is: 
\begin{align}
	J^\mu(t, \vec{x}) &= J^\mu_\textsf{rad}(t, \vec{x}) + J^\mu_\textsf{stat}(\vec{x})\notag\\
	&= {e}^{-i \omega t} J^\mu_{0}(\vec{x})+ J^\mu_\textsf{stat}(\vec{x}).
\end{align}

  By inserting this ansatz into the formal solution~\eqref{eq:formal_solution}, we obtain
\begin{align}\label{eq:ansatz_A}
	A^\mu(t, \vec{x}) &= \int_\Omega\d^3 x'\, J^\mu_0(\vec{x}') \frac{{e}^{ i\omega \|\vec{x}-\vec{x}'\|}}{\|\vec{x}-\vec{x}'\|}{e}^{-i\omega t} + A^\mu_\textsf{stat}(\vec{x}),
\end{align}
where $A^\mu_\textsf{stat}(\vec{x})$ contains the static contributions.

Finally, we must take into account the position of the observer relative to the source, which can be organised into three different zones:
\begin{enumerate}
	\item The near zone: $d\ll r\ll \lambda$
	\item The transition zone: $d\ll r \simeq \lambda$
	\item The far/radiation zone: $d\ll \lambda \ll r$
\end{enumerate}
The behavior of the vector potential is  different in the three zones and this directly impacts the observer's ability to locally infer the existence of electromagnetic radiation. 

Let us analyse the near and the far zone. In the former case, the condition $r\ll \lambda$ allows us to expand the exponential in~\eqref{eq:ansatz_A} and we find, by also applying an expansion of~\eqref{eq:ansatz_A} in spherical harmonics,
\begin{align}
	A^\mu(t, \vec{x}) &= \sum_{l=0}^\infty\sum_{|m|\leq l}\frac{{e}^{-i\omega t}}{2l+1}\frac{Y_{lm}(\theta,\phi)}{r^{l+1}}\int_\Omega\d^3 x' J^\mu_0(\vec{x}')\, r'^{l} Y^*_{lm}(\theta',\phi') + A^\mu_\textsf{stat}(\vec{x}) & \text{(for $r\ll \lambda$)}.
\end{align}
Although this expression \emph{is time-dependent},  the dependence is not the one expected of radiation, since the oscillation does not depend on the distance to the source. Fields that oscillate in this way are called \textbf{quasi-static}~\citep{JacksonBook}. Moreover, the fall-off is not $\frac{1}{r}$, but rather a sum over terms with  $\frac{1}{r^{l+1}}$ as coefficients. Again, this  is not the behavior of a radiation field, and so  an observer would not be able to unambiguously infer the existence of electromagnetic waves in this region.

In the far zone, we implement the condition $\lambda\ll r$ and expand $\|\vec{x}-\vec{x}'\|$ as
\begin{equation}
	\|\vec{x}-\vec{x}'\| \approx r - \vec{n}\cdot \vec{x}',
\end{equation} 
where $\vec{n}$ is a unit vector in the direction of $\vec{x}$. Using this approximation in \eqref{eq:ansatz_A}, we obtain:
\begin{equation}\label{eq:FarZone}
	\lim_{r\to \text{far zone}} A^\mu(\vec{x}, t) = \frac{1}{4\pi} \frac{e^{i \omega (r-t)}}{r}\int_\Omega\d^3 x'\, J^\mu_0(\vec{x}')\,{e}^{-i \omega\vec{n}\cdot \vec{x}'} + A^\mu_\textsf{stat}(\vec{x}).
\end{equation}
The first term in the above expression has the expected properties: it oscillates, it decays like $\frac{1}{r}$, and it is transversal: this is a genuine radiation field.

The conclusion of this argument is that the observer has to be far enough away from the source to effectively detect radiation. Too close, and the vector potential is quasi-static. 

One could attempt to characterise radiation by the energy and momentum that they carry. But while the flux carries energy and momentum, a non-zero flux may have no associated electromagnetic radiation. As mentioned in the main text, the Poynting vector alone cannot parse  radiation and other field contributions, except asymptotically. 

For consider the Poynting vector $\vec{S}:=\vec{E}\times\vec{B}$ with its associated Poynting flux, $\oint_{\bbS^2}\vec{S}\, \d^2\sigma$, where $\bbS^2$ is a $2$-sphere and $\d^2\sigma = r^2\, \sin\theta\,\d\theta\,\d\phi$.  The Poynting vector is not a Lorentz invariant quantity and it therefore depends on a choice of reference frame. As an example, consider the Coulomb solution, i.e., the field of a point charge for an observer in the rest frame of the particle. Clearly, for such an observer the magnetic field is zero and consequently the Poynting vector vanishes. 
But a boosted observer will see a current, rather than a static charge, and thus an electric and a magnetic field, and so:
\begin{equation}
	\vec{S}_\textsf{rest frame} = 0 \neq \vec{S}_\textsf{boosted}.
\end{equation}

But the Poynting flux \emph{carries} information about electromagnetic radiation, in the asymptotic limit. An explicit computation for the above example shows that the boosted observer sees a Poynting vector which decays like $\frac{1}{r^4}$, and thus the Poynting flux of the boosted observer vanishes at infinity. That is, we obtain
\begin{equation}
	\lim_{r\to\infty}\int_{\bbS^2} \vec{S}_\textsf{rest frame}\,\d^2\sigma = 0 = \lim_{r\to\infty}\int_{\bbS^2} \vec{S}_\textsf{boosted}\,\d^2\sigma.
\end{equation}
Both observers now agree that there is no electromagnetic radiation. Going to infinity filters all but the radiative parts of a field. More precisely,  the Poynting flux of static contributions vanishes at infinity while the Poynting flux of electromagnetic waves is non-zero.


The standard way to disentangle different components according to their fall-off conditions is to use a conformal compactification of asymptotically flat spacetimes, and a Penrose-Newman null tetrad decomposition, to which we now turn. (See \cite[Ch. 3,4]{Dambrosio_grav}.

 \section{Gravitational radiation}\label{app:grav}

Let  $(\hat{M}, \hat{g}_{ab})$ be a physical spacetime which satisfies Einstein's field equations with vanishing cosmological constant, $\hat{R}_{ab}-\frac12 \hat{R}\,\hat{g}_{ab} = 8\pi\, \hat{T}_{ab}$.\footnote{Hatted quantities are the physical quantities, so why encumber notation with hats? Because in this subject, the corresponding asymptotically completed notions are more used, and so they are honoured with the unhatted symbols.} We call this spacetime \textbf{asymptotically Minkowski} if it satisfies the three following conditions
\begin{itemize}
	\item[1)] There exists a conformal completion $(M, g_{ab}, \Omega)$ such that $M := \hat{M}\, \cup\mathscr{I}$ is a manifold with a boundary and the boundary has the topology $\mathscr{I}\simeq \mathbb{S}^2\times \RR$. Moreover, the conformally rescaled metric and the physical metric are related by $g_{ab} = \Omega^2\,\hat{g}_{ab}$. The conformal factor is assumed to satisfy $\Omega\,\hat{=}\,0$ and $\nabla_a\Omega\,\hat{\neq}\, 0$, where hatted equalities are equalities on $\mathscr{I}$.
	\item[2)] $\Omega^{-2}\hat{T}_{ab}$ has a smooth limit to $\mathscr{I}$. 
	\item [3)] The normal vector field to $\mathscr{I}$, $n^{a}:=\left.g^{ab}\nabla_b\Omega\right|_{\Omega=0}$, is complete.
\end{itemize}
This definition still allows full conformal freedom at $\mathscr{I}$; a canonical partial fixing of this freedom is given by a \textbf{divergence-free conformal frame}, for which $\nabla_a n^{a}\, \hat{=}\, 0$. The choice of such a frame still allows a conformal rescaling on $\mathbb{S}^2$ (that is dragged along by $n$). Moreover, using the asymptotic limit for the Einstein equations in the conformally completed spacetime, we obtain:
\be \nabla_a n_b\,\hat{=}\,0\ee 

The conditions $\Omega\,\hat{=}\,0$ and $\nabla_a \Omega\,\hat{\neq}\, 0$ tell us that $\Omega$ is a good coordinate near $\mathscr{I}$, that $\mathscr{I}$ has a well-defined normal $n^{a} := \left.\nabla^a\Omega\right|_{\Omega=0}$, and that $\Omega$ is heuristically the same as $\frac{1}{r}$.\footnote{Were we to take a stronger convergence for the conformal completion, say $\Omega=\frac{1}{r^2}$, then we wouldn't obtain a null generator from the conformal factor, since $\nabla_a\Omega\,\hat{=}\,0$.}  Thus the condition that $\Omega^{-2}\,\hat{T}_{ab}$ has a smooth limit to $\mathscr{I}$  tells us that $\hat{T}_{ab}$ falls-off  \textit{at least} like $\frac{1}{r^2}$. One finds that this is a condition which is satisfied by all reasonable compact sources.

The null tetrad is given, in the physical metric, by null vectors $\hat{n}, \hat{\ell}, \hat{m}, \hat{\bar m}$, satisfying the following conditions: 
\be  \hat {n}^a\hat{\ell}_a=-1,\quad \hat {m}^a\hat{\bar m}_a=1, 
\ee
and all other inner products vanishing. This definition implies we can write the metric as:
\be \hat{g}_{ab}=-2\hat{n}_{(a}\hat{\ell}_{b)}+2\hat{m}_{(a}\hat{\bar m}_{b)}.
\ee
Given a conformal compactification, there exists a corresponding conformally rescaled null tetrad $\ell^a$ and $m, \bar m$ that is well defined at $\mathscr{I}$ (with $\ell^a=\Omega^{-2}\hat{\ell}^a$ and $m^a=\Omega^{-1}\hat{m}^a$).\footnote{In fact, it is convenient to go in the opposite direction: defining a null tetrad in $\mathscr{I}^+$ and then dragging it back into the bulk (see \cite[Sec. 3.C]{Dambrosio_grav}). First, we chose $n^{a}$ as the first null normal to $\mathscr{I}$. This vector field is defined on all of $\mathscr{I}$ and it allows us to introduce an affine parameter $u$ which foliates $\mathscr{I}$. We have then introduced $(\theta, \phi)$ coordinates on the $u=$ const. leaves of the foliation. Put together, $(u, \theta, \phi)$ provides us with a globally defined coordinate system for $\mathscr{I}$. Next, we have introduced a Newman-Penrose null tetrad $\{\mathring{\ell}^{a}, \mathring{n}^{a}, \mathring{m}^{a}, \mathring{\bar{m}}^{a}\}$ on a cross-section $\mathring{\mathbb{S}}^2$. This ``reference'' null tetrad is normalized in the usual way and it serves as ``generator'' of a null tetrad on all of $\mathscr{I}$. In fact, we can generate such a null tetrad by Lie dragging (or parallel transporting, which is the same in this context) the reference tetrad along $n^{a}$ (or along its integral lines). Finally, we have extended the null tetrad from $\mathscr{I}$ into a neighborhood of $\mathscr{I}$ by Lie dragging it along $\ell^{a}$ into the bulk of spacetime. }

And we can then use these to decompose the electromagnetic field tensor as: 
\begin{align}
	\Phi_2 &:=  F_{ab}  n^{a} \bar{m}^b\notag\\
	\Phi_1 &:= \frac12  F_{ab}\left( n^{a} \ell^{b} +  m^{a}\bar{m}^{b}\right)\notag\\
	\Phi_0 &:=  F_{ab}  m^{a}  \ell^{b}.
\end{align}
Given the relationship between $\Omega$ and $1/r$, and the relationship between the physical tetrads and the conformally completed one, we obtain a Peeling theorem for the electromagnetic tensor: a definite fall-off rate for each of the scalars above. Tthe  only component that falls-off as $1/r$ is $\Phi_2$: that is the component that we asymptotically associate with radiation.  

In the gravitational case, we apply a similar treatment to the Weyl curvature.  
 Although, unlike the Faraday tensor, the Weyl curvature vanishes at $\mathscr{I}$ (see \cite[Sec. 3.D]{Dambrosio_grav}), one applies the decomposition to the conformally rescaled Weyl tensor, $K_{abcd}=\Omega^{-1}C_{abcd}$:
 \begin{align}
	{\Psi}_4 &:= K_{abcd}\,n^{a} \bar{m}^{b} n^{c} m^{d}\notag\\
	{\Psi}_3 &:= K_{abcd}\,\ell^{a} n^{b} \bar{m}^c n^d\notag\\
	{\Psi}_2 &:= K_{abcd}\,\ell^{a} m^{b} \bar{m}^c n^d\notag\\
	{\Psi}_1 &:= K_{abcd}\,\ell^{a} n^{b} \ell^{c} m^{d}\notag\\
	{\Psi}_0 &:= K_{abcd}\, \ell^{a} m^{b}\ell^{c} m^{d} 
\end{align}
From this we can find again a Peeling theorem, that leads again, just as in the case of electromagnetism, to a neat separation of the different components. We find that $ \Psi_4$  encodes the radiation field, since it decays like $\frac{1}{r}$; and $\Psi_2$ encodes the ``Coulombic'' information of the gravitational field (i.e., the mass of the source which generates the field).

That is, denoting the limit of the Weyl scalars at $\mathscr{I}$ by $\Psi^\circ$, one can write:
\be\label{eq:Psi_shear}  \Psi_4^\circ = -\ddot{\bar{\sigma}}^\circ \ee
where $\sigma^\circ$ is the asymptotic shear,
defined as
\begin{equation}
   \label{eq:sigma_Dell} \sigma^\circ(u,\theta,\phi) = - \lim_{r\to\infty} \left(m^{a}m^{b}\nabla_a \ell_b\right),
\end{equation}
where we have extended all fields into the bulk of spacetime.\footnote{Here we have avoided explicitly discussing the intrinsic geometry of $\mathscr{I}$, which is degenerate, with a degenerate metric $q_{ab}$---that is non-degenerate on  the two dimensional hypersurfaces (isomorphic to) $\mathbb{S}^2$ foliating $\mathscr{I}$, and degenerate along $u$, or $n$ (so its inverse is ambiguous, and all results must be shown to be independent of this ambiguity). In fact, the shear tensor can be described by $q_{ab}$ and the intrinsic covariant derivative on $\mathscr{I}$, which we label $\mathcal{D}$, as the trace-free:
\begin{align}\label{eq:RunningOutOfNames}
    \sigma_{ab}^\circ := -\mathcal{D}_a \ell_b + \frac12 q_{ab} q^{cd} (\mathcal{D}_c \ell_d).
\end{align}
It is the trace-freeness that guarantees that the end result is fully conformally invariant. Since the shear tensor $\sigma_{ab}$ is transverse, trace-less, and symmetric (that is, it satisfies $\sigma_{ab} n^{b} = 0$, $q^{ab}\sigma_{ab}$, and $\sigma_{[ab]} = 0$) this implies that the shear is of the form
\begin{align*}
    \sigma_{ab} = -(\bar{\sigma}^\circ m_a m_b + \sigma^\circ \bar{m}_a \bar{m}_b).
\end{align*}
From this equation we obtain \eqref{eq:sigma_Dell}.}

 This limit can be explicitly computed and compared with the linearized theory, where $h_+$ and $h_\times$ encode the radiative modes. In this linearised regime we find
\begin{equation}
    \sigma^\circ(u,\theta,\phi) = \frac12\left(h^\circ_+ + i\, h^\circ_\times\right)(u, \theta, \phi).
\end{equation}
We refer to $h^\circ_+$ and $h^\circ_\times$ as the \textbf{strains} of the gravitational wave where we defined
\begin{align}
    h^\circ_+(u,\theta,\phi) &:= \lim_{r\to\infty}r h_+ (u,r,\theta,\phi),\notag\\ 
    h^\circ_\times(u,\theta,\phi) &:= \lim_{r\to\infty}r h_\times(u,r,\theta,\phi).
\end{align}
Thus we find:
\begin{align}\label{eq:Psi4inStrain}
    \Psi^\circ_4=-\frac12 \left(\Ddot{h}_+^\circ -i\Ddot{h}_\times^\circ\right),
\end{align}

Thus, we have established a connection between the Newman-Penrose scalar $\Psi^\circ_4$ and the strains of the gravitational wave we use in the linearized theory. The label ``radiation field'' is thus well-justified for $\Psi^\circ_4$. It is worth remarking that this links theory to observations and data analysis. In fact, $\Psi^\circ_4$ is a key quantity which is computed in Numerical Relativity and integrating it twice over $\d u$, isolates the strains. This is what is ultimately used in waveform models and plotted in the famous waveform plots.

There is, of course, much more to be said, about conservation laws---which refer to the symmetry group of $\mathscr{I}$---and, equally important,  about how $\sigma^\circ(u,\theta,\phi)$ (and its complex conjugate) encodes the entire  conformal geometry of $\mathscr{I}$, with a vanishing shear corresponding to no radiation. 

Indeed, the Bondi News tensor is constructed entirely from geometric properties of $\mathscr{I}$: namely, it is uniquely determined by  the trace-free part of the Schouten tensor at  $\mathscr{I}$. And it is related to the asymptotic shear as follows:
\begin{align}\label{eq:bondiI}
    N_{ab} = 2\mathcal{L}_n \sigma_{ab}^\circ =: \dot{\sigma}_{ab}^\circ.
\end{align}
In terms of conservation laws, the energy flux across a portion of $\mathscr{I}$ can  be expressed as
\be\label{eq:GWFluxes}
    F_E(\Delta\mathscr{I}) = \int_{\Delta\mathscr{I}} N_{ab} N_{cd} q^{ca} q^{bd}\, \d u\,\d^2\omega\ee
 
This is as much as we can fit into this small appendix.

 

\end{document}